\documentclass[reprint,superscriptaddress,amsmath,amssymb,aps]{revtex4-1}
\usepackage[utf8]{inputenc}
\usepackage[T1]{fontenc} 
\usepackage{graphicx}
\usepackage{rotating}
\usepackage{makeidx}
\usepackage{threeparttable}
\usepackage{float}
\usepackage{amsmath}
\usepackage{rotating}
\usepackage{hyperref}
\usepackage{siunitx}
\usepackage{relsize}
\usepackage{etoolbox}
\usepackage[most]{tcolorbox}
\usepackage{hyperref}
\usepackage{dcolumn}
\usepackage{bm}
\usepackage{wrapfig}
\usepackage{float}  
\usepackage[toc,page]{appendix}
\usepackage{gensymb}
\usepackage{epstopdf}
\usepackage{subfiles}
\usepackage{comment} 
\usepackage[autostyle=true]{csquotes} 
\usepackage{makecell}
\usepackage{siunitx}
\usepackage[all]{hypcap} 
\usepackage[most]{tcolorbox}
\usepackage{hyperref}
\usepackage{xcolor}
\usepackage{nicefrac}
\usepackage{chemformula}
\usepackage{ulem}  
\definecolor{gainsboro}{rgb}{0.86, 0.86, 0.86}
\usepackage[backgroundcolor=gainsboro]{todonotes}
\hypersetup{
  colorlinks   = true, 
  urlcolor     = blue, 
  linkcolor    = blue, 
  citecolor   = blue 
}

\newcommand{\weg}[1]{}

\newcommand{\obyth}[2]{$\frac{1}{3}$}
\newcommand{\obyf}[2]{$\frac{1}{4}$}

\begin{document}

\title{Ballistic photocurrent driven by optical
phonon modes in a polaronic ferroelectric}

\author{Sangeeta Rajpurohit}
\email{srajpurohit@lbl.gov}
\affiliation{Molecular Foundry, Lawrence Berkeley
National Laboratory, USA}

\author{Tadashi Ogitsu}
\affiliation{Lawrence Livermore National Laboratory,
Livermore, USA}

\author{Liang Z. Tan}
\affiliation{Molecular Foundry,
Lawrence Berkeley National Laboratory, USA}
\date{\today}

\begin{abstract}
We investigate the effect of local electron-phonon
coupling on nonlinear optical conductivity in an
interacting ferroelectric system. Using real-time
simulations, we show an enhancement in nonlinear
conductivity under linearly-polarized light due to
generation of the phonon-assisted ballistic-current
in addition to the injection-current generated
by electron-hole pairs. The optically excited phonon
modes generate an asymmetric carrier distribution that
causes a strong directional ballistic-current. The
ballistic-current enhances the photocurrent several
times at above band-gap excitation frequencies and is
sublinearly dependent on the excitation intensity. This
strong phonon-assisted zero-frequency directional ballistic-current
demonstrates an alternative way to boost the bulk
photovoltaic effect (BPVE) in electronic ferroelectric
materials with strong local el-ph coupling.  
\end{abstract}

\maketitle

The nonlinear optical effects in quantum
materials have been extensively studied due to
their strong fundamental and technological relevance \cite{Bloembergen1996,Boyd2020}.
For example, the coupling between the polarization
and optical properties in solids lacking inversion
symmetry, such as ferroelectrics, causes unconventional
photovoltaic effects, known as bulk photovoltaic effect (BPVE) \cite{Baltz1981,Belinicher1980}.
The BPVE is the generation of directional (dc)
photocurrent due to above-bandgap electronic
excitations under an external electric
field  \cite{Baltz1981,Kraut1979}.

The dc-current generated in BPVE can be
classified into two types: shift-current
and ballistic-current \cite{Sturman1992,Sturman2020,Tan2016}.
The shift-current is a purely coherent
quantum phenomenon that involves the relative
shift of the electron cloud in real space
during excitation under linearly polarized
light. It is independent of any scattering
mechanism, and its relaxation occurs at
electronic timescales. On the other hand, the
ballistic-current (also referred to as injection current)
arises from the asymmetric momentum distribution of
carriers due to scattering processes such as electron-phonon
(el-ph)~\cite{Belinicher1978, Belinicher1980, Belinicher1988, Sturman1992, Burger2019, Menshenin2003},
electron-electron (el-el)~\cite{Kaneko2021}, and
electron-spin (el-s) scattering. Another form of
injection current is present under circularly polarized
light, where population asymmetry is generated by
photoexcitation instead of scattering~\cite{Sipe2000}.  
Higher-order photocurrents which involve population
asymmetry depend on various kinetic processes that
span a range of time scales \cite{Mahon2019,Kral2000},
unlike the coherent shift current process.

While past theoretical studies of the
BPVE have mostly been performed in perturbation
theory, recent works have shown that naive application
of perturbation theory can lead to incorrect
results for the shift and injection currents
in certain limits~\cite{Matsyshyn2021}.
Furthermore, the use of the independent-particle
approximation in the frozen band picture
limits the predictive power of such studies
in non-perturbation regions \cite{Kaneko2021}.
Dynamical effects, such as carrier momentum
and energy relaxation due to elastic and inelastic
scattering, are known to significantly affect the
BPVE properties of materials \cite{Barik2020,Matsyshyn2021,Kaneko2021,Rajpurohit2021}.
The rate of relaxation of energy and momentum 
depends on the interactions el-ph, el-el, and el-s 
and impurities, all of which are intrinsic
properties of materials \cite{Rajpurohit2020, Sotoudeh2017}. In addition to
carrier scattering, quasi-particle effects, such
as excitonic dynamics, are also known to affect
photocurrent dynamics and produce a non-trivial
photovoltaic response \cite{Barik2020,Matsyshyn2021,Kaneko2021,Rajpurohit2021}.

The phonon-induced ballistic current has only
been treated within perturbation theory in
previous studies~\cite{Belinicher1978,Dai2021}.
With the limitations of perturbation theory
for shift and injection currents in mind,
we reexamine the phonon-induced ballistic
current using non-pertubative real-time
simulations in the present study. We
demonstrate that a phonon-induced ballistic
current  can be sustained at steady state,
using a one-dimensional polaronic ferroelectric
system as an illustrative model. 
Importantly, local el-ph coupling
(of the Holstein-type) favors the generation
of this photocurrent. These optical phonon
modes coupled to charge-transfer electronic
excitations cause strong asymmetric el-ph
scattering that generates a ballistic photocurrent.
This phonon-assisted current is unidirectional,
and results in a strong BPVE response
independent of light polarization. The magnitude
of the reported ballistic-current originating
from optical phonon modes is the largest contributor
to the BPVE in this system. We observe a departure
from the predictions perturbation theory in the
high intensity limit. Our reported large contribution of
the ballistic-current to BPVE originating
from phonon dynamics can explain the 
temperature dependence of the photocurrent, 
which cannot be resolved by the shift current
theory, recently observed in the ferroelectric
organic molecular system \cite{Nakamura2017}. 

Our study suggests possibilities for
strengthening the nonlinear optical rectification
and the BPVE effect in ferroelectrics with strong
el-ph interactions displaying bound excitations 
and long-lived optically excited high-frequency
phonon modes, such as complex transition-metal
oxides or charge-transfer organic salts.

We consider a minimal one-dimensional ferroelectric model
with an el-ph interaction. Motivated by three-dimensional
ABO$_3$ transition-metal perovksite oxides, the model
consists of a chain of corner connected BO$_6$-octahedra
with the $z$-axis along the chain. The Hilbert space
consists of a single electron orbital on each B-type ion.
Electronic hopping between these B-type orbitals is
mediated via intermediate O-type sites, similar to
transition metal oxides. These electrons are described by
one-particle wavefunctions where each wave function
is a two-component spinor. The wavefunctions are
expressed as $|\psi_n\rangle=\sum_{j}|\chi_{\sigma,j}\rangle\psi_{\sigma,j,n}$
with band index $n$ and occupation $f_n$ in terms of
local spin-orbital $|\chi_{\sigma,j}\rangle$ at every
$j^{th}$ B-type site. The potential energy $E_{pot}(|\psi_n\rangle,Q_{j})$ in 
this interacting ferroelectric model is defined as 
\begin{eqnarray}
E_{pot} &=&\sum_nf_n\sum_j(t_{hop}+(-1)^{j}\delta)(\psi_{\sigma,j,n}\psi_{\sigma,j+1,n}^* \nonumber \\
&+&\psi_{\sigma,j+1,n}\psi_{\sigma,j,n}^*) -g\sum_{\sigma,j}\rho_{\sigma,\sigma,j}Q_{j} +K/2\sum_{j}Q^2_{j} \nonumber \\ 
&+&U/2\sum_{j,\sigma\ne\sigma'}(\rho_{\sigma,\sigma,j}\rho_{\sigma',\sigma',j}-\rho_{\sigma,\sigma',j}\rho_{\sigma',\sigma,j})
\label{eqn:model}
\end{eqnarray}
The first and second terms correspond to the electron
hopping between B-sites and the el-ph interaction,
respectively. $t_{hop}$ and $g$ are the strengths of
the hopping and el-ph coupling. For every $j^{th}$ B-site,
we consider the local octahedral breathing mode $Q_{j}$,
defined as $Q_{j}{=}1/\sqrt(3)(d_{x,i}+d_{y,i}+d_{z,i}-3\bar{d})$ 
where $d_{x/y/z,i}$ is the distance between O-type ions
forming octahedra around the $i^{th}$ B-site along $x/y/z$
direction and $\bar{d}=3.85$ \r{A} is the equilibrium O-O
distance. These $Q_{j}$ modes are coupled to the electron
density at the respective $j^{th}$ B-sites. The third term
in $E_{pot}$ is the potential energy of the displacements
of $u_i$, where $K$ is the restoring force constant.

\begin{figure}[tp!]
     \begin{center}
     \includegraphics[width=\linewidth]{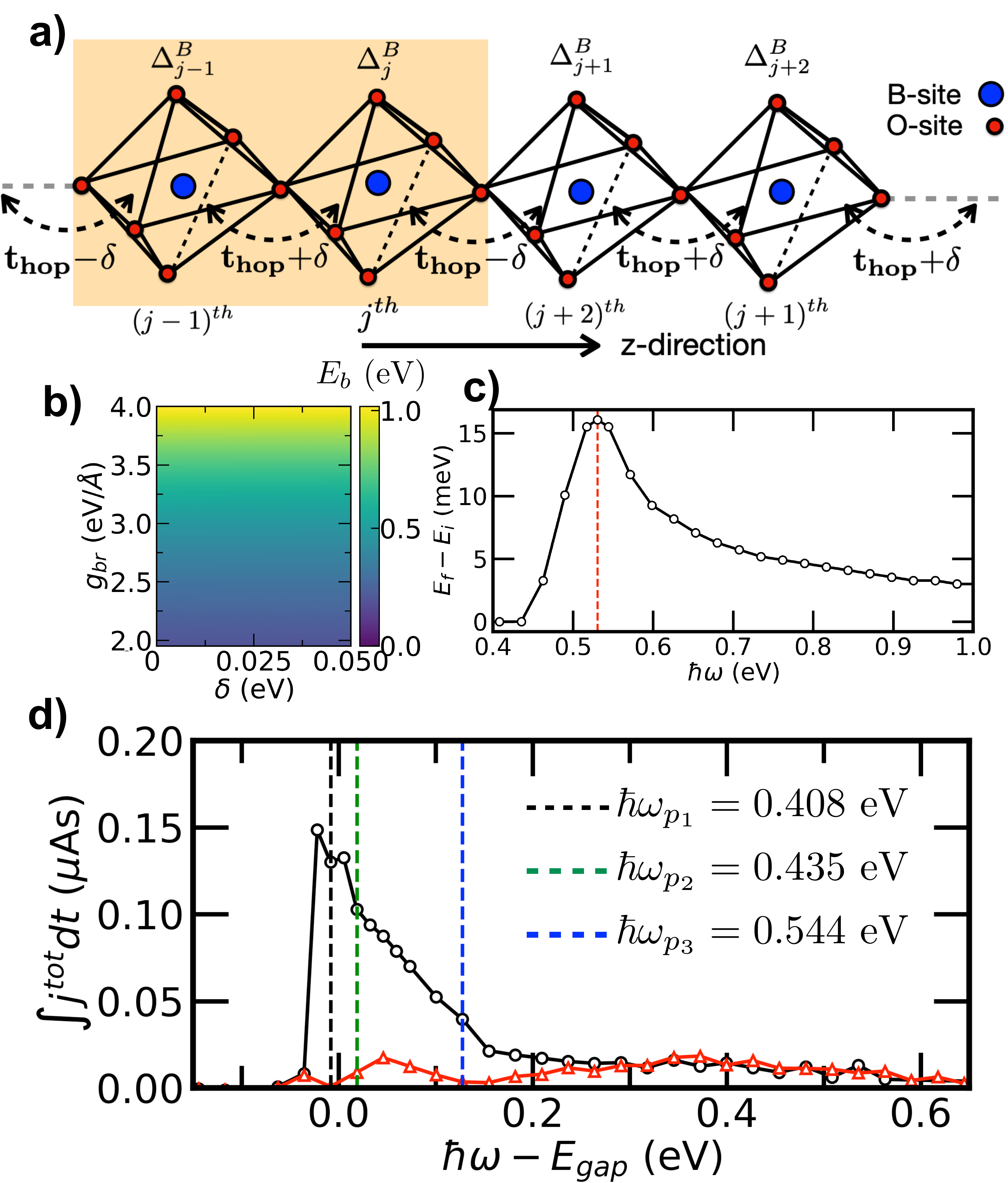}
     \caption{a): One dimensional chain of corner-connected
     BO$_6$-octahedra where B-type cations (blue circles) and O-type
     anions (red circles) occupy the center and corner of octahedra.
     The unit cell is shown in yellow; b): Polaron binding energy $E_b=(E_{pot}-E^{|Q=0|}_{pot})$, 
     within $g-\delta$. c): Photon absorption density as
     a function of excitation energy. d): integrated current
     $\int j^{tot}(t) dt$ as a function of excitation energy. The
     lines in black and red show $\int j^{tot}(t) dt$ in the presence and absence
     of atom dynamics, respectively. b),c) and d) consider $t_{hop}=0.50$ eV,
     $U=0.50$ eV, K=10 eV/\r{A}$^2$. c) and d) considers $\delta=0.025$ eV and
      $g{=}2.80$ eV/\r{A}. d) uses $I=3.14$ $\times 10^4$ W/cm$^2$
      (i.e., $A_o=0.001$ $\hbar/ea_o$). }
     \label{fig:fig1}
     \end{center}
\end{figure}

The restoring force constant $K$ is related to the mass $M_O$ and the
vibration frequency of the phonon mode $Q_i$ by $K{=}M_O\omega_b^2$.
We fix $K{=}10$ eV/\r{A}$^2$ so that the vibration frequency of the
$Q_j$ modes, assuming $M_O$ as the mass of oxygen atoms, is in the
range of the frequencies of optical phonon modes such as Jahn-Teller
and breathing modes, in transition-metal oxides. 

Finally, we include local el-el interactions via the
$U$ parameter. This term incorporates a fast decoherence
and relaxation effect within the electronic subsystem.
These effects are present in most materials, and serve
to bring the system to a steady state under photoexcitation.

We calculate the ground state phase diagram of the model,
defined in Eqn \ref{eqn:model}, at half-filling (one electron
per BO$_6$-site) as a  function of the model parameters
($g,\delta$), with the phononic ($Q$) and electronic ($\psi$)
degrees of freedom allowed to relax to their lowest energy
configuration. Finite $Q_{j}$ modes act as a onsite potential
$\Delta^B_j=gQ_{j}$ for electrons. For $U=0$, the ground-state
exhibits an insulating phase with a charge density wave (CDW)
for $g>g_c$. The CDW is accompanied by lattice distortions
reflected in finite $Q_{j}$ modes. An increase in the onsite
$U$ drives the system toward the spin density wave (SDW).
This is in agreement with previous theoretical studies of the 
two-site Hubbard-Holstein model \cite{Fehske2003}.

In the present work, we focus on the CDW  insulating phase.
Figure \ref{fig:fig1}-b shows the magnitude of $Q_i$ in the
$g$-$\delta$ plane. The ground states obtained with
$\delta\ne0$ and $\Delta^B_i\ne 0$ lack inversion symmetry
and are ferroelectric -- these are equivalent to the
Rice-Mele 1d-ferroelectric model that is often used
to study the effects of BPVE \cite{Rice1982}. The real-time
study with $U{\ne}0$ is a time-dependent version of the
interacting Rice-Mele model. As we focus on the role of
the el-ph interaction $g_{br}$ in BPVE, we keep $U/t_{hop}=1$ constant
throughout our study, so that the system is always in
the CDW insulating phase.

The ferroelectric phase favors the formation of polarons.
The binding energy of the polaron $E_b$ is equal to the
energy difference between the distorted case $Q_i{\ne}0$
and the undistorted case $Q_i{=}0$. Figures \ref{fig:fig1}-b
show the variation of $E_b$ in the $g_{br}$-$\delta$ plane. 
The typical polaron binding energies measured in hole-doped
transition-metal oxides are in the range of 100-500 meV \cite{Mildner2015}.
In our current 1-d model, the range of parameters
$g_{br}=2.50{-}3.50$ eV/\r{A} and $\delta=0-0.05$ eV
reproduces a similar $E_b$.

To simulate the real-time dynamics of photocurrent generation
and its evolution under a light field, we employ Ehrenfest dynamics. 
The one-particle electron wavefunctions evolve under
the time-dependent Schrodinger equation,  while the
atoms obey the classical equations of motion.

The effect of the linearly polarized light field, defined
by the vector field $\vec{A}=\Big(A_oe^{\iota\omega_ot}{+}A_oe^{-\iota\omega_o t}\Big)\vec{e}_{z}$, 
where $A_o$ is the amplitude of the vector potential
and $\hbar\omega_o$ is the photon energy, is incorporated
into the model \ref{eqn:model} using the Peierls substitution
method \cite{Peierls1933}.

For the real-time simulations of the BPVE, we consider the
polaronic ferroelectric state as the initial state with
band-gap $E_{gap}$=0.411 eV at time $t{=}0$. In the initial
state, the chain is dimerized with a staggered pattern
of local distortion $Q_i$, and with zero initial velocities.
The BPVE effect is investigated in the parameter ranges
$\delta=0.0-0.025$ eV and $g{=}2.75-3.0$ eV/\r{A} while
keeping $t_{hop}=0.50$ eV and $U=0.50$ eV fixed.

The simulations are carried out with a time step $dt{=}0.96 \times 10^{-17}$
s in a supercell with 4 A-type sites and  periodic boundary conditions.
We used a $N_k$=800 point k-grid symmetric around the $\Gamma$-point. 
\begin{figure}[tp!]
     \begin{center}
     \includegraphics[width=\linewidth]{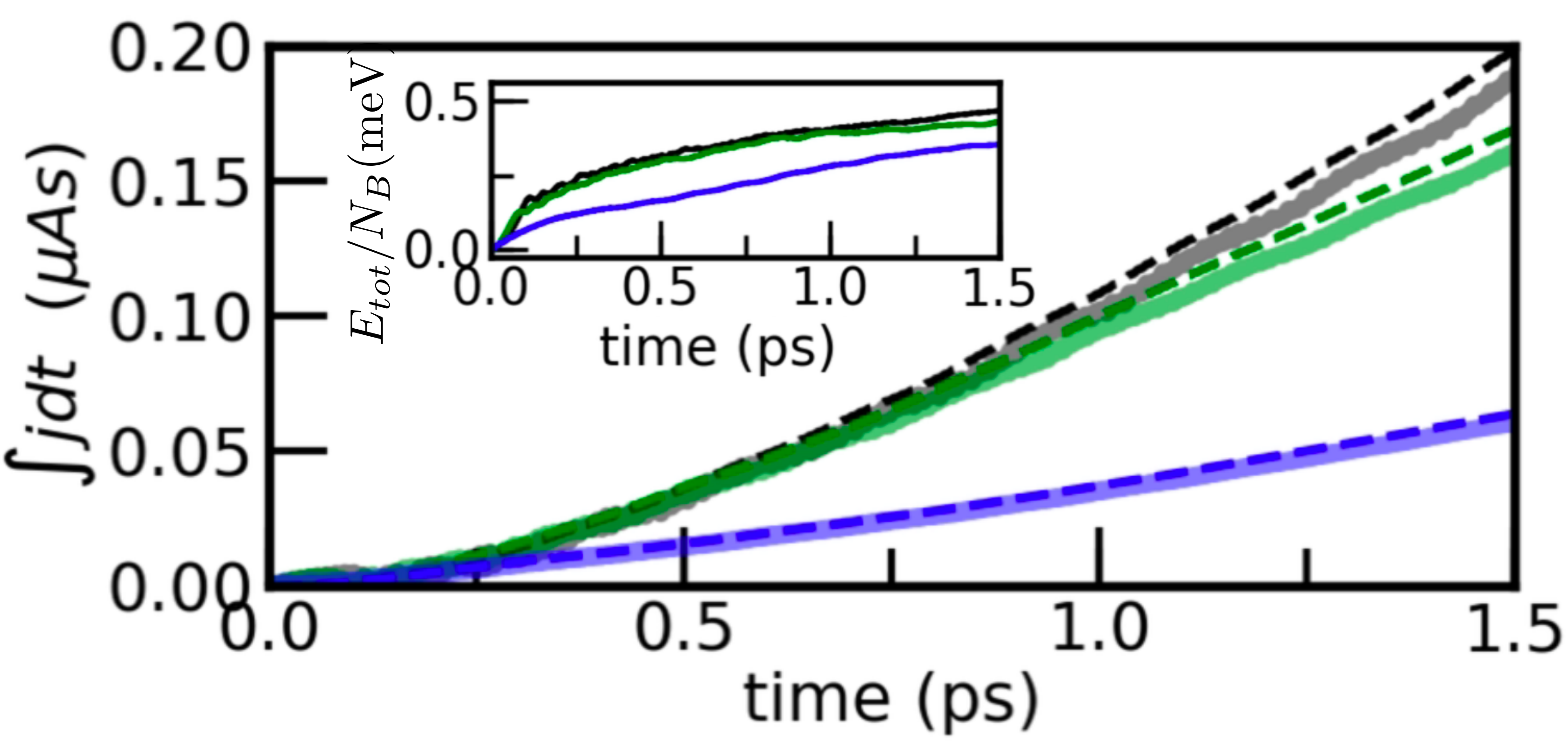}
     \caption{a): Evolution of total current
     $\int j^{tot}(t,\omega_p) dt$ (dashed-lines) and $\int j^{scattering}(t) dt$
     current component (faint-colored lines) at photon energy $\hbar\omega_{p_1}=0.408$ eV (black),
     $\hbar\omega_{p_2}=0.435$ eV (green) and $\hbar\omega_{p_3}=0.544$ eV (blue), as
     indicated in Fig \ref{fig:fig1}-d. Inset shows the evolution of the total energy $E_{tot}=E_{kin}+E_{pot}$.}
     \label{fig:fig2}
     \end{center}
\end{figure}
Firstly, we investigate the spectral distribution the photon absorption
density $D_p{=}\delta E_{f-i}/{\hbar\omega}$, where $\delta E_{f-i}$
is a change in $E_{tot}=E_{Kin}+E_{pot}$, before and after a short
20-femtosecond Gaussian-shaped light pulse. The system shows a broad
absorption peak with a band-gap of $0.411$ eV, see Figure \ref{fig:fig1}-c. 

Next, we consider the effect of a continuous-wave light field on
the photocurrent. Figures \ref{fig:fig1}-d shows the spectrum
of the time integrated-current $\int^{t_f}_{t_i} j^{tot}(t)dt$. 
where the total current $j^{tot}(t) $ is defined as 
\begin{eqnarray}
 j^{tot}(t)&=&\sum_nf_n\sum_je^{i\vec{A}(t)\bar{d}}(t_{hop}+(-1)^{j}\delta) \nonumber \\
&&(\psi_{j,n}(t)\psi_{j+1,n}^*(t) -\psi_{j+1,n}(t)\psi_{j,n}^*(t)) \Big) \vec{e}_{z}.
   \label{eq:charge_current}
\end{eqnarray}
The current is integrated over a period $t_i{=}$1.08 ps to $t_f{=}$1.45 ps.
In the presence of atom dynamics, the system displays a strong photocurrent
over a wide energy range between $\hbar\omega=0.40-0.60$ eV. We attribute
this photocurrent to ballistic-current $j^{ballistic}(t)$ generated as
a result of asymmetric carrier scattering by phonon modes. In comparison,
the photocurrent in the frozen-atom case is several times smaller. 
We attribute this contribution to the photocurrent to injection current 
induced by electron-hole (el-h) pair creation under linearly polarized 
light as discussed in Ref.~\cite{Kaneko2021}.

\begin{figure}[tp!]
     \begin{center}
     \includegraphics[width=\linewidth]{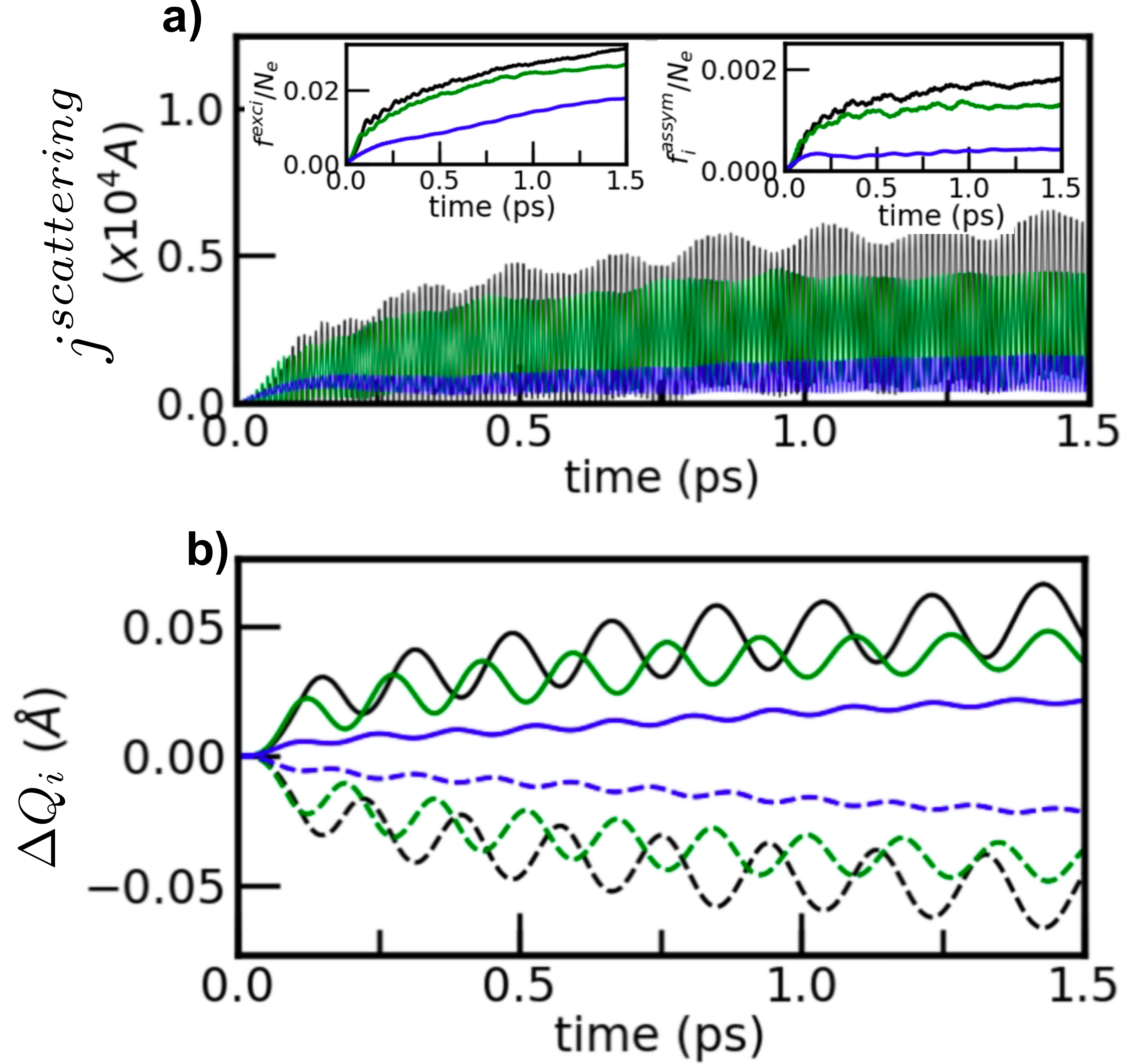}
     \caption{a): Evolution of instantaneous current $\int j^{scattering}(t) dt$
     current. Left-inset shows fraction of excited electrons. 
     Right-inset show evolution asymmetric carrier momentum distribution 
     $f^{assym}=\sum_{v,k}(\tilde{f}_{v,k}{-}\tilde{f}_{v,-k})$ of valence band.
     b) dynamics of local phonon modes $\Delta Q^B_i=Q^B_i(t)-Q^B_i(t=0)$,
     where Negative and positive $\Delta Q_i(t)$ corresponds to
     different B-type sites of the unit cell. For description of the colors, 
     see Fig \ref{fig:fig2}. }
     \label{fig:fig3}
     \end{center}
\end{figure}

In figures \ref{fig:fig2} and \ref{fig:fig3}, we show the evolution
of the total photocurrent and carrier populations as a function
of time at three different excitation frequencies $\hbar\omega_p$.
In the beginning, the dynamics is highly non-equilibrium in nature,
with a continuous increase in the photocurrent, excited-state
populations and total energy $E_{tot}=E_{pot}+E_{kin}$ of
the system, where $E_{kin}$ is the kinetic energy of atoms.
For excitation at $\omega_p{_1}$, the system reaches quasi
steady-state at time $t{\sim}1.0$ ps, which is reflected in
the saturation of the photocurrent, excited state populations,
and total energies. The time taken to reach steady state
increases as the excitation frequency is increased to
$\omega_p{_2}$ and $\omega_p{_3}$. The correlation between
the saturation of the photocurrent with the saturation of
the carrier populations is consistent with the photocurrent
being mostly ballistic current, which is sensitive to the
carrier populations.

Optical excitations from the valence band to the conduction
band in the polaronic ferroelectric state are accompanied by
charge-transfer between sites, altering local charge
densities $\rho_i(t)$. The phonon modes $Q_i(t)$
are coupled with local change densities $\rho_i(t)$
Thus, changes in $\rho_i(t)$ induce atomic displacements,
which are primarily high-frequency optical phonon modes,
as seen in Figures \ref{fig:fig3}-b.   

Bulk photocurrents arise in the scattering picture~\cite{Dai2020}
when carriers occupy an asymmetric carrier momentum distribution
$\tilde{f}_{m,k}\ne \tilde{f}_{m,-k}$ in the Brillouin zone.
This asymmetric carrier momentum distribution can be induced by
scattering from phonons or el-h pairs. Thus, the sum
$j^{scattering}(t)$ of the ballistic- and injection-current
induced by el-ph and el-hole scattering is  
\begin{eqnarray}
j^{scattering}(t) &=&j^{ballistic} (t) + j^{inj}(t) \nonumber \\
 &=&-e\sum_{k,m}\tilde{f}_{m}(k,t) v_{m}(k,t)
\label{eqn:sh_b}
\end{eqnarray}
where $\tilde{f}_{m}(k,t)$ and $v_m(k,t)$
are the momentum distribution and the electron
velocity in band $m$, respectively.
For the present two-band model, the electron
velocities $v_{m}(k,t)$ in band m are
$v_{m}(k,t)=\partial_k\epsilon_{m}(k,t)$. 
We compute $v_{m}(k,t)$ using a finite difference method,
\begin{eqnarray}
v_{m}(k,t)=\frac{\epsilon_{m}(k+\delta k,t)-\epsilon_{m,t}(k-\delta k,t)}{2\delta k}
\label{eq:ballis}
\end{eqnarray}

The occupancy $\tilde{f}_{m}(k,t)$ of the eigenstates
$\phi_{m}^{BO}(k,t)\rangle$ of the instantaneous electronic
Hamiltonian is calculated by projecting the occupied
one-particle wavefunctions $|\psi_n(k,t)\rangle$ on these
eigenstates. 

 In general, asymmetric velocities $v_{m,k}\ne v_{m,-k}$ or
by asymmetric occupations $\tilde{f}_{m,k}\ne \tilde{f}_{m,-k}$
can both give rise to bulk photocurrents, see Eqn \ref{eq:ballis}.
In this model, the generation of ballistic- and injection-current
is entirely due to asymmetric occupations because time-reversal
symmetry of the model forbids asymmetric band velocities.

On comparing the computed scattering current $j^{scattering}(t)$
with the total current $j^{tot}(t)$ (Figure~\ref{fig:fig2}-a),
we find that most of the photocurrent in this system can be
explained by scattering current. The remainder, which arises
from carrier coherences instead of carrier populations~\cite{Tan2016},
is understood to be the shift current. In this one-dimensional polaronic 
ferroelectric insulator, electronic transitions from the
valence band to the conduction band are accompanied by
a shift in the charge center which constitutes
the shift current $j^{sh}(t)$. We find that the shift
current $j^{sh}(t)=j^{tot}(t){-}j^{scattering}(t)$, contribution
remains very small in the presence and absence of atom dynamics.

Figures \ref{fig:fig3}-a show the evolution of the
instantaneous current $j^{scattering}(t)$,
defined in Eqn \ref{eqn:sh_b}, for different values of
$\hbar\omega_p$.  The evolution of the asymmetric carrier
momentum distribution $\sum_{v,k} (\tilde{f}_{v,k}{-}\tilde{f}_{v,-k})$ 
in the valence band is shown in Figure \ref{fig:fig3}-a right-inset. Like
the total current calculated from Eq.~\ref{eq:charge_current}, assymetric
carrier momentum distribution also becomes almost
constant for $t>0.75$ ps, showing that a constant asymmetric
carrier scattering rate is achieved at steady state.
Even though the el-el interaction $U$ is small enough that el-ph scattering
accounts for most of the photocurrent, el-el scattering plays a
role in bringing the system to a steady state by acting as a
relaxation process that limits asymmetric carrier distributions.

\begin{figure}[tp!]
     \begin{center}
     \includegraphics[width=\linewidth]{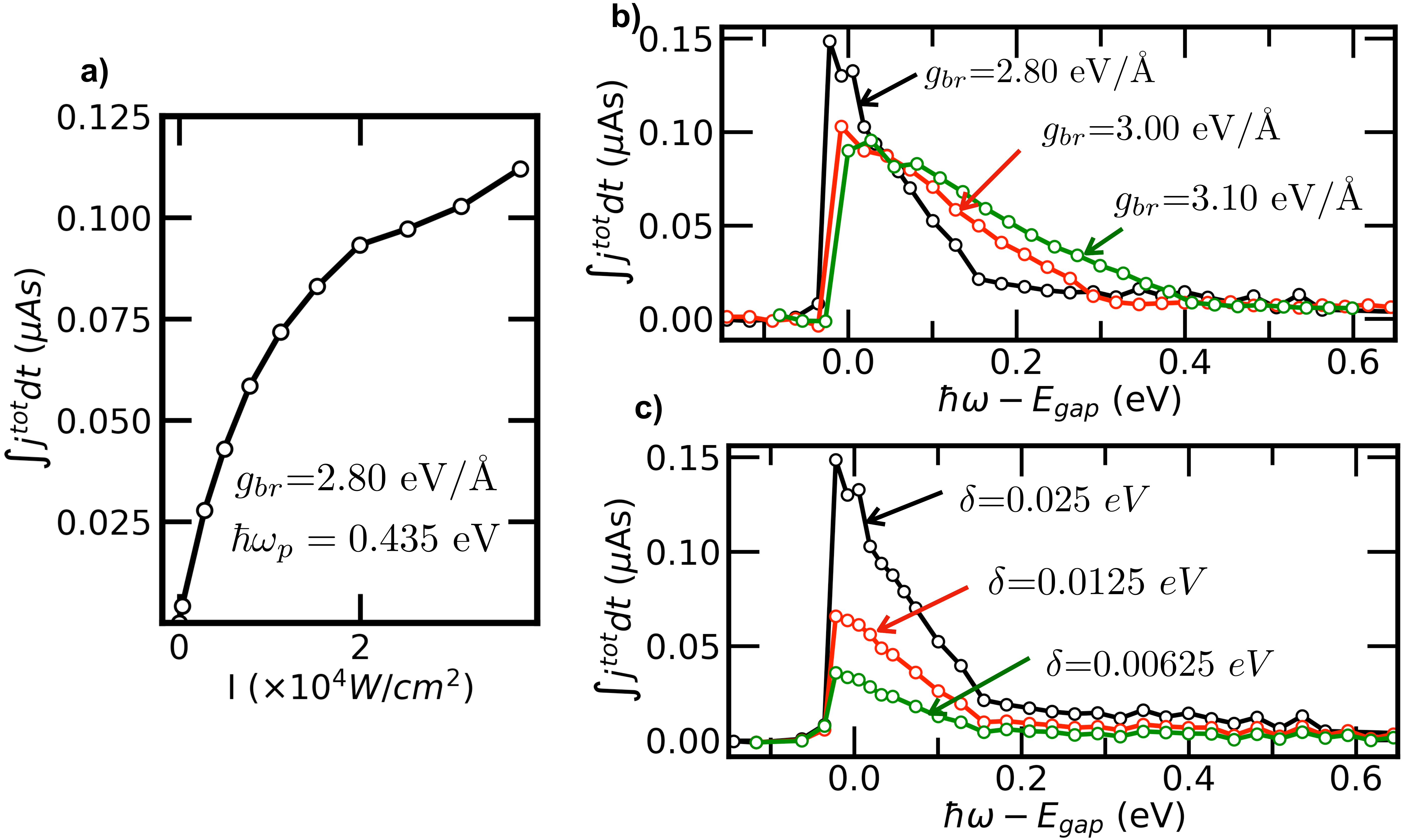}
     \caption{a) Integrated-current $\int j^{tot}(t)dt$ as a function
     of excitation intensity showing a sublinear behavior.
     b) $\int j^{tot}(t)dt$ for different values
     of el-ph coupling strength $g_{br}$; c) $\int j^{tot}(t)dt$
     for different values of $\delta$. Figure a) 
     uses $g_{br}=2.80$ eV/\r{A} and $\delta=0.05$ eV. 
     Figure b) uses  $I=5.0$ mW (i.e.,  $A_o=0.001$ $\hbar/ea_o$)
     and $\delta=0.025$ eV. Figure c) uses $I=3.14$ $\times 10^4$ W/cm$^2$ 
     (i.e., $A_o=0.001$ $\hbar/ea_o$) and $g_{br}=2.80$ eV/\r{A}.}
     \label{fig:fig4}
     \end{center}
\end{figure}

The strength of these relaxation processes in comparison to the
light intensity $I=1/2|A_o^2|\omega_o^2 c\epsilon_o$ is an important parameter that controls the behavior
of bulk photocurrents. Figure \ref{fig:fig4}-a shows the behavior
of the total current with intensity in the presence of atom dynamics. 
We observe a crossover from a linear current dependence on intensity
to a deviation from linear behavior at higher intensities. The
low-intensity limit is consistent with the predictions of
perturbation theory~\cite{Dai2021}.  A crossover to the high
intensity limit was also obtained for injection currents under
circularly polarized light using the Keldysh-Floquet
formalism~\cite{Matsyshyn2021}, where it was predicted that
the current scales as the square root of intensity in the high
intensity (slow relaxation) limit. Our numerical simulations
suggest that the  Keldysh-Floquet analytical results derived
for injection currents under circularly polarized light are
likely to apply to other bulk photocurrents such as the
phonon-induced ballistic photocurrent.

The magnitude of the phonon-induced ballistic current depends
strongly on the form of el-ph coupling. In this model, the
Holstein-type el-ph coupling, where the electron density couples
directly to the phonon modes ($\rho_jQ_j$, Eq.~\ref{eqn:model}),
is responsible for creating a strongly asymmetric carrier
distribution. Polaronic materials, which include many ferroelectric
materials, are likely to have an el-ph coupling containing a
term of this form. In comparison, Peierls-like el-ph coupling \cite{Kaneko2021},
in which phonon modes modify the hopping amplitudes between sites,
is expected to result in very small ballistic currents induced
by phonons. Unlike this non-local el-ph coupling, our study
suggests that the local el-ph coupling may have a pronounced
impact on BPVE.

Now we consider the effect of the
el-ph coupling parameter $g$ and the hopping asymmetry $\delta$
on the current. Figure \ref{fig:fig4}-b shows the total current
$\int j^{tot}(t)dt$ for four different el-ph coupling strengths.
Even though some amount of el-ph coupling is necessary to create
an asymmetric carrier distribution, a larger value of el-ph coupling
does not always result in larger ballistic current. For this range
of parameters, we observe that $\int j^{tot}(t)dt$ decreases
with increasing $g_{br}$ close to the band edge, but has the
opposite trend away from the band edge. This is because besides
asymmetric carrier scattering rate, the carrier velocities have a
significant impact on the ballistic current (Eq.~\ref{eqn:sh_b}).
The renormalization of the carrier velocities near the band edge
is reflected in the higher band edge $\int j^{tot}(t)dt$ at
$g=2.80$ eV/\r{A} compared to $g=$3.00 eV/\r{A} and 3.10 eV/\r{A}. 

The model parameter $\delta$, see Eqn. 1, describes the relative
asymmetry between the bonds. A higher value of $\delta$ means
greater asymmetry between the bonds. Changing $\delta$ from
$0.00625$ eV to $0.0125$ and $0.025$ eV (Figure \ref{fig:fig4}-c )
enhances the current almost by a factor of two and four,
respectively. This shows that a higher $\delta$ generates
a stronger BPVE. This is similar to the behavior of the
shift-current, which also increases with $\delta$.

Our present study highlights the importance of optical phonon
modes and the polaronic effect in inducing a strong photocurrent
in ferroelectric insulators. We show that Holstein-type el-ph
coupling produces strong zero-frequency ballistic-current.
Multiferroic oxides and charge-transfer organic salts
are potential candidates for the realization of strong BPVE.
In multiferroic oxides, ferroelectricity originates from
the underlying charge and orbital order that are strongly
coupled to local atomic distortions, and the band gap
is sensitive to these distortions \cite{Sotoudeh2017,Rajpurohit2021}.
The d-d-type electronic transitions across the bandgap
in such systems are expected to generate a strong non-linear
photovoltaic response.

Using a tight-binding model with realistic parameters, we
have shown that phonon-induced ballistic current can be
supported in a polaronic ferroelectrics, and it can be
the dominant contributor to the bulk photocurrent when
the right form of el-ph coupling is present. The high-frequency
phonon modes that are coupled to electronic excitations generate
an asymmetric momentum distribution of charge carriers
and induce a strong directional ballistic photocurrent.
Our study suggests that the effects of optical phonon
modes strongly influence the BPVE properties, and
understanding the contribution of these effects is
essential to boost the bulk photovoltaic response in
ferroelectrics known to exhibit
polaronic character.

S.R., T.O., and L.Z.T were supported by the Computational
Materials Sciences Program funded by the US Department
of Energy, Office of Science, Basic Energy Sciences,
Materials Sciences and Engineering Division. Additional
support for absorption calculations was obtained from
the Molecular Foundry, a DOE Office of Science User Facility
supported by the Office of Science of the U.S. Department
of Energy under Contract No. DE-AC02-05CH11231.
\bibliography{ref}

\begin{thebibliography}{28}%
\makeatletter
\providecommand \@ifxundefined [1]{%
 \@ifx{#1\undefined}
}%
\providecommand \@ifnum [1]{%
 \ifnum #1\expandafter \@firstoftwo
 \else \expandafter \@secondoftwo
 \fi
}%
\providecommand \@ifx [1]{%
 \ifx #1\expandafter \@firstoftwo
 \else \expandafter \@secondoftwo
 \fi
}%
\providecommand \natexlab [1]{#1}%
\providecommand \enquote  [1]{``#1''}%
\providecommand \bibnamefont  [1]{#1}%
\providecommand \bibfnamefont [1]{#1}%
\providecommand \citenamefont [1]{#1}%
\providecommand \href@noop [0]{\@secondoftwo}%
\providecommand \href [0]{\begingroup \@sanitize@url \@href}%
\providecommand \@href[1]{\@@startlink{#1}\@@href}%
\providecommand \@@href[1]{\endgroup#1\@@endlink}%
\providecommand \@sanitize@url [0]{\catcode `\\12\catcode `\$12\catcode
  `\&12\catcode `\#12\catcode `\^12\catcode `\_12\catcode `\%12\relax}%
\providecommand \@@startlink[1]{}%
\providecommand \@@endlink[0]{}%
\providecommand \url  [0]{\begingroup\@sanitize@url \@url }%
\providecommand \@url [1]{\endgroup\@href {#1}{\urlprefix }}%
\providecommand \urlprefix  [0]{URL }%
\providecommand \Eprint [0]{\href }%
\providecommand \doibase [0]{http://dx.doi.org/}%
\providecommand \selectlanguage [0]{\@gobble}%
\providecommand \bibinfo  [0]{\@secondoftwo}%
\providecommand \bibfield  [0]{\@secondoftwo}%
\providecommand \translation [1]{[#1]}%
\providecommand \BibitemOpen [0]{}%
\providecommand \bibitemStop [0]{}%
\providecommand \bibitemNoStop [0]{.\EOS\space}%
\providecommand \EOS [0]{\spacefactor3000\relax}%
\providecommand \BibitemShut  [1]{\csname bibitem#1\endcsname}%
\let\auto@bib@innerbib\@empty
\bibitem [{\citenamefont {Bloembergen}(1996)}]{Bloembergen1996}%
  \BibitemOpen
  \bibfield  {author} {\bibinfo {author} {\bibfnamefont {N.}~\bibnamefont
  {Bloembergen}},\ }\href {\doibase 10.1142/3046} {\emph {\bibinfo {title}
  {Nonlinear Optics}}},\ \bibinfo {edition} {4th}\ ed.\ (\bibinfo  {publisher}
  {WORLD SCIENTIFIC},\ \bibinfo {year} {1996})\ \Eprint
  {http://arxiv.org/abs/https://www.worldscientific.com/doi/pdf/10.1142/3046}
  {https://www.worldscientific.com/doi/pdf/10.1142/3046} \BibitemShut {NoStop}%
\bibitem [{\citenamefont {Boyd}(2020)}]{Boyd2020}%
  \BibitemOpen
  \bibfield  {author} {\bibinfo {author} {\bibfnamefont {R.}~\bibnamefont
  {Boyd}},\ }\href {\doibase 10.1142/3046} {\emph {\bibinfo {title} {Nonlinear
  Optics}}},\ \bibinfo {edition} {4th}\ ed.\ (\bibinfo  {publisher} {Academic
  Press},\ \bibinfo {year} {2020})\BibitemShut {NoStop}%
\bibitem [{\citenamefont {von Baltz}\ and\ \citenamefont
  {Kraut}(1981)}]{Baltz1981}%
  \BibitemOpen
  \bibfield  {author} {\bibinfo {author} {\bibfnamefont {R.}~\bibnamefont {von
  Baltz}}\ and\ \bibinfo {author} {\bibfnamefont {W.}~\bibnamefont {Kraut}},\
  }\href {\doibase 10.1103/PhysRevB.23.5590} {\bibfield  {journal} {\bibinfo
  {journal} {Phys. Rev. B}\ }\textbf {\bibinfo {volume} {23}},\ \bibinfo
  {pages} {5590} (\bibinfo {year} {1981})}\BibitemShut {NoStop}%
\bibitem [{\citenamefont {Belinicher}\ and\ \citenamefont
  {Sturman}(1980)}]{Belinicher1980}%
  \BibitemOpen
  \bibfield  {author} {\bibinfo {author} {\bibfnamefont {V.~I.}\ \bibnamefont
  {Belinicher}}\ and\ \bibinfo {author} {\bibfnamefont {B.~I.}\ \bibnamefont
  {Sturman}},\ }\href {\doibase 10.1070/pu1980v023n03abeh004703} {\bibfield
  {journal} {\bibinfo  {journal} {Soviet Physics Uspekhi}\ }\textbf {\bibinfo
  {volume} {23}},\ \bibinfo {pages} {199} (\bibinfo {year} {1980})}\BibitemShut
  {NoStop}%
\bibitem [{\citenamefont {Kraut}\ and\ \citenamefont {von
  Baltz}(1979)}]{Kraut1979}%
  \BibitemOpen
  \bibfield  {author} {\bibinfo {author} {\bibfnamefont {W.}~\bibnamefont
  {Kraut}}\ and\ \bibinfo {author} {\bibfnamefont {R.}~\bibnamefont {von
  Baltz}},\ }\href {\doibase 10.1103/PhysRevB.19.1548} {\bibfield  {journal}
  {\bibinfo  {journal} {Phys. Rev. B}\ }\textbf {\bibinfo {volume} {19}},\
  \bibinfo {pages} {1548} (\bibinfo {year} {1979})}\BibitemShut {NoStop}%
\bibitem [{\citenamefont {Sturman}\ and\ \citenamefont
  {Fridkin}(1992)}]{Sturman1992}%
  \BibitemOpen
  \bibfield  {author} {\bibinfo {author} {\bibfnamefont {B.~I.}\ \bibnamefont
  {Sturman}}\ and\ \bibinfo {author} {\bibfnamefont {V.~M.}\ \bibnamefont
  {Fridkin}},\ }\href@noop {} {\emph {\bibinfo {title} {The Photovoltaic and
  Photorefractive Effects in Noncentrosymmetric Materials}}}\ (\bibinfo
  {publisher} {Gordon and Breach Science Publishers},\ \bibinfo {year}
  {1992})\BibitemShut {NoStop}%
\bibitem [{\citenamefont {Sturman}(2020)}]{Sturman2020}%
  \BibitemOpen
  \bibfield  {author} {\bibinfo {author} {\bibfnamefont {B.~I.}\ \bibnamefont
  {Sturman}},\ }\href {\doibase 10.3367/ufne.2019.06.038578} {\bibfield
  {journal} {\bibinfo  {journal} {Physics-Uspekhi}\ }\textbf {\bibinfo {volume}
  {63}},\ \bibinfo {pages} {407} (\bibinfo {year} {2020})}\BibitemShut
  {NoStop}%
\bibitem [{\citenamefont {Tan}\ \emph {et~al.}(2016)\citenamefont {Tan},
  \citenamefont {Zheng}, \citenamefont {Young}, \citenamefont {Wang},
  \citenamefont {Liu},\ and\ \citenamefont {Rappe}}]{Tan2016}%
  \BibitemOpen
  \bibfield  {author} {\bibinfo {author} {\bibfnamefont {L.~Z.}\ \bibnamefont
  {Tan}}, \bibinfo {author} {\bibfnamefont {F.}~\bibnamefont {Zheng}}, \bibinfo
  {author} {\bibfnamefont {S.~M.}\ \bibnamefont {Young}}, \bibinfo {author}
  {\bibfnamefont {F.}~\bibnamefont {Wang}}, \bibinfo {author} {\bibfnamefont
  {S.}~\bibnamefont {Liu}}, \ and\ \bibinfo {author} {\bibfnamefont {A.~M.}\
  \bibnamefont {Rappe}},\ }\href {\doibase 10.1038/npjcompumats.2016.26}
  {\bibfield  {journal} {\bibinfo  {journal} {npj Computational Materials}\
  }\textbf {\bibinfo {volume} {2}},\ \bibinfo {pages} {16026} (\bibinfo {year}
  {2016})}\BibitemShut {NoStop}%
\bibitem [{\citenamefont {Belinicher}\ and\ \citenamefont
  {Sturman}(1978)}]{Belinicher1978}%
  \BibitemOpen
  \bibfield  {author} {\bibinfo {author} {\bibfnamefont {V.~I.}\ \bibnamefont
  {Belinicher}}\ and\ \bibinfo {author} {\bibfnamefont {B.~I.}\ \bibnamefont
  {Sturman}},\ }\href@noop {} {\bibfield  {journal} {\bibinfo  {journal} {Sov.
  Phys. Solid State}\ }\textbf {\bibinfo {volume} {20}},\ \bibinfo {pages}
  {476} (\bibinfo {year} {1978})}\BibitemShut {NoStop}%
\bibitem [{\citenamefont {Belinicher}\ and\ \citenamefont
  {Sturman}(1988)}]{Belinicher1988}%
  \BibitemOpen
  \bibfield  {author} {\bibinfo {author} {\bibfnamefont {V.~I.}\ \bibnamefont
  {Belinicher}}\ and\ \bibinfo {author} {\bibfnamefont {B.~I.}\ \bibnamefont
  {Sturman}},\ }\href {\doibase 10.1080/00150198808235446} {\bibfield
  {journal} {\bibinfo  {journal} {Ferroelectrics}\ }\textbf {\bibinfo {volume}
  {83}},\ \bibinfo {pages} {29} (\bibinfo {year} {1988})},\ \Eprint
  {http://arxiv.org/abs/https://doi.org/10.1080/00150198808235446}
  {https://doi.org/10.1080/00150198808235446} \BibitemShut {NoStop}%
\bibitem [{\citenamefont {Burger}\ \emph {et~al.}(2019)\citenamefont {Burger},
  \citenamefont {Agarwal}, \citenamefont {Aprelev}, \citenamefont {Schruba},
  \citenamefont {Gutierrez-Perez}, \citenamefont {Fridkin},\ and\ \citenamefont
  {Spanier}}]{Burger2019}%
  \BibitemOpen
  \bibfield  {author} {\bibinfo {author} {\bibfnamefont {A.~M.}\ \bibnamefont
  {Burger}}, \bibinfo {author} {\bibfnamefont {R.}~\bibnamefont {Agarwal}},
  \bibinfo {author} {\bibfnamefont {A.}~\bibnamefont {Aprelev}}, \bibinfo
  {author} {\bibfnamefont {E.}~\bibnamefont {Schruba}}, \bibinfo {author}
  {\bibfnamefont {A.}~\bibnamefont {Gutierrez-Perez}}, \bibinfo {author}
  {\bibfnamefont {V.~M.}\ \bibnamefont {Fridkin}}, \ and\ \bibinfo {author}
  {\bibfnamefont {J.~E.}\ \bibnamefont {Spanier}},\ }\href {\doibase
  10.1126/sciadv.aau5588} {\bibfield  {journal} {\bibinfo  {journal} {Science
  Advances}\ }\textbf {\bibinfo {volume} {5}},\ \bibinfo {pages} {eaau5588}
  (\bibinfo {year} {2019})}\BibitemShut {NoStop}%
\bibitem [{\citenamefont {Men’shenin}(2003)}]{Menshenin2003}%
  \BibitemOpen
  \bibfield  {author} {\bibinfo {author} {\bibfnamefont {V.~V.}\ \bibnamefont
  {Men’shenin}},\ }\href {\doibase 10.1134/1.1626750} {\bibfield  {journal}
  {\bibinfo  {journal} {Physics of the Solid State}\ }\textbf {\bibinfo
  {volume} {45}},\ \bibinfo {pages} {2131} (\bibinfo {year}
  {2003})}\BibitemShut {NoStop}%
\bibitem [{\citenamefont {Kaneko}\ \emph {et~al.}(2021)\citenamefont {Kaneko},
  \citenamefont {Sun}, \citenamefont {Murakami}, \citenamefont
  {Gole\ifmmode~\check{z}\else \v{z}\fi{}},\ and\ \citenamefont
  {Millis}}]{Kaneko2021}%
  \BibitemOpen
  \bibfield  {author} {\bibinfo {author} {\bibfnamefont {T.}~\bibnamefont
  {Kaneko}}, \bibinfo {author} {\bibfnamefont {Z.}~\bibnamefont {Sun}},
  \bibinfo {author} {\bibfnamefont {Y.}~\bibnamefont {Murakami}}, \bibinfo
  {author} {\bibfnamefont {D.}~\bibnamefont {Gole\ifmmode~\check{z}\else
  \v{z}\fi{}}}, \ and\ \bibinfo {author} {\bibfnamefont {A.~J.}\ \bibnamefont
  {Millis}},\ }\href {\doibase 10.1103/PhysRevLett.127.127402} {\bibfield
  {journal} {\bibinfo  {journal} {Phys. Rev. Lett.}\ }\textbf {\bibinfo
  {volume} {127}},\ \bibinfo {pages} {127402} (\bibinfo {year}
  {2021})}\BibitemShut {NoStop}%
\bibitem [{\citenamefont {Sipe}\ and\ \citenamefont
  {Shkrebtii}(2000)}]{Sipe2000}%
  \BibitemOpen
  \bibfield  {author} {\bibinfo {author} {\bibfnamefont {J.~E.}\ \bibnamefont
  {Sipe}}\ and\ \bibinfo {author} {\bibfnamefont {A.~I.}\ \bibnamefont
  {Shkrebtii}},\ }\href {\doibase 10.1103/PhysRevB.61.5337} {\bibfield
  {journal} {\bibinfo  {journal} {Phys. Rev. B}\ }\textbf {\bibinfo {volume}
  {61}},\ \bibinfo {pages} {5337} (\bibinfo {year} {2000})}\BibitemShut
  {NoStop}%
\bibitem [{\citenamefont {Mahon}\ \emph {et~al.}(2019)\citenamefont {Mahon},
  \citenamefont {Muniz},\ and\ \citenamefont {Sipe}}]{Mahon2019}%
  \BibitemOpen
  \bibfield  {author} {\bibinfo {author} {\bibfnamefont {P.~T.}\ \bibnamefont
  {Mahon}}, \bibinfo {author} {\bibfnamefont {R.~A.}\ \bibnamefont {Muniz}}, \
  and\ \bibinfo {author} {\bibfnamefont {J.~E.}\ \bibnamefont {Sipe}},\ }\href
  {\doibase 10.1103/PhysRevB.100.075203} {\bibfield  {journal} {\bibinfo
  {journal} {Physical Review B}\ }\textbf {\bibinfo {volume} {100}},\ \bibinfo
  {pages} {075203} (\bibinfo {year} {2019})},\ \bibinfo {note} {publisher:
  American Physical Society}\BibitemShut {NoStop}%
\bibitem [{\citenamefont {Král}\ and\ \citenamefont {Sipe}(2000)}]{Kral2000}%
  \BibitemOpen
  \bibfield  {author} {\bibinfo {author} {\bibfnamefont {P.}~\bibnamefont
  {Král}}\ and\ \bibinfo {author} {\bibfnamefont {J.~E.}\ \bibnamefont
  {Sipe}},\ }\href {\doibase 10.1103/PhysRevB.61.5381} {\bibfield  {journal}
  {\bibinfo  {journal} {Physical Review B}\ }\textbf {\bibinfo {volume} {61}},\
  \bibinfo {pages} {5381} (\bibinfo {year} {2000})}\BibitemShut {NoStop}%
\bibitem [{\citenamefont {Matsyshyn}\ \emph {et~al.}(2021)\citenamefont
  {Matsyshyn}, \citenamefont {Piazza}, \citenamefont {Moessner},\ and\
  \citenamefont {Sodemann}}]{Matsyshyn2021}%
  \BibitemOpen
  \bibfield  {author} {\bibinfo {author} {\bibfnamefont {O.}~\bibnamefont
  {Matsyshyn}}, \bibinfo {author} {\bibfnamefont {F.}~\bibnamefont {Piazza}},
  \bibinfo {author} {\bibfnamefont {R.}~\bibnamefont {Moessner}}, \ and\
  \bibinfo {author} {\bibfnamefont {I.}~\bibnamefont {Sodemann}},\ }\href
  {\doibase 10.1103/PhysRevLett.127.126604} {\bibfield  {journal} {\bibinfo
  {journal} {Phys. Rev. Lett.}\ }\textbf {\bibinfo {volume} {127}},\ \bibinfo
  {pages} {126604} (\bibinfo {year} {2021})}\BibitemShut {NoStop}%
\bibitem [{\citenamefont {Barik}\ and\ \citenamefont {Sau}(2020)}]{Barik2020}%
  \BibitemOpen
  \bibfield  {author} {\bibinfo {author} {\bibfnamefont {T.}~\bibnamefont
  {Barik}}\ and\ \bibinfo {author} {\bibfnamefont {J.~D.}\ \bibnamefont
  {Sau}},\ }\href {\doibase 10.1103/PhysRevB.101.045201} {\bibfield  {journal}
  {\bibinfo  {journal} {Physical Review B}\ }\textbf {\bibinfo {volume}
  {101}},\ \bibinfo {pages} {045201} (\bibinfo {year} {2020})},\ \bibinfo
  {note} {publisher: American Physical Society}\BibitemShut {NoStop}%
\bibitem [{\citenamefont {{Rajpurohit}}\ \emph {et~al.}(2021)\citenamefont
  {{Rajpurohit}}, \citenamefont {{Das Pemmaraju}}, \citenamefont {{Ogitsu}},\
  and\ \citenamefont {{Tan}}}]{Rajpurohit2021}%
  \BibitemOpen
  \bibfield  {author} {\bibinfo {author} {\bibfnamefont {S.}~\bibnamefont
  {{Rajpurohit}}}, \bibinfo {author} {\bibfnamefont {C.}~\bibnamefont {{Das
  Pemmaraju}}}, \bibinfo {author} {\bibfnamefont {T.}~\bibnamefont {{Ogitsu}}},
  \ and\ \bibinfo {author} {\bibfnamefont {L.~Z.}\ \bibnamefont {{Tan}}},\
  }\href@noop {} {\bibfield  {journal} {\bibinfo  {journal} {arXiv e-prints}\
  ,\ \bibinfo {eid} {arXiv:2105.11310}} (\bibinfo {year} {2021})},\ \Eprint
  {http://arxiv.org/abs/2105.11310} {arXiv:2105.11310 [cond-mat.str-el]}
  \BibitemShut {NoStop}%
\bibitem [{\citenamefont {{Rajpurohit}}\ \emph {et~al.}(2020)\citenamefont
  {{Rajpurohit}}, \citenamefont {{Jooss}},\ and\ \citenamefont
  {{Bl{\"o}chl}}}]{Rajpurohit2020}%
  \BibitemOpen
  \bibfield  {author} {\bibinfo {author} {\bibfnamefont {S.}~\bibnamefont
  {{Rajpurohit}}}, \bibinfo {author} {\bibfnamefont {C.}~\bibnamefont
  {{Jooss}}}, \ and\ \bibinfo {author} {\bibfnamefont {P.~E.}\ \bibnamefont
  {{Bl{\"o}chl}}},\ }\href {\doibase 10.1103/PhysRevB.102.014302} {\bibfield
  {journal} {\bibinfo  {journal} {Phys. Rev. B}\ }\textbf {\bibinfo {volume}
  {102}},\ \bibinfo {pages} {014302} (\bibinfo {year} {2020})}\BibitemShut
  {NoStop}%
\bibitem [{\citenamefont {Sotoudeh}\ \emph {et~al.}(2017)\citenamefont
  {Sotoudeh}, \citenamefont {Rajpurohit}, \citenamefont {Bl\"ochl},
  \citenamefont {Mierwaldt}, \citenamefont {Norpoth}, \citenamefont {Roddatis},
  \citenamefont {Mildner}, \citenamefont {Kressdorf}, \citenamefont {Ifland},\
  and\ \citenamefont {Jooss}}]{Sotoudeh2017}%
  \BibitemOpen
  \bibfield  {author} {\bibinfo {author} {\bibfnamefont {M.}~\bibnamefont
  {Sotoudeh}}, \bibinfo {author} {\bibfnamefont {S.}~\bibnamefont
  {Rajpurohit}}, \bibinfo {author} {\bibfnamefont {P.}~\bibnamefont
  {Bl\"ochl}}, \bibinfo {author} {\bibfnamefont {D.}~\bibnamefont {Mierwaldt}},
  \bibinfo {author} {\bibfnamefont {J.}~\bibnamefont {Norpoth}}, \bibinfo
  {author} {\bibfnamefont {V.}~\bibnamefont {Roddatis}}, \bibinfo {author}
  {\bibfnamefont {S.}~\bibnamefont {Mildner}}, \bibinfo {author} {\bibfnamefont
  {B.}~\bibnamefont {Kressdorf}}, \bibinfo {author} {\bibfnamefont
  {B.}~\bibnamefont {Ifland}}, \ and\ \bibinfo {author} {\bibfnamefont
  {C.}~\bibnamefont {Jooss}},\ }\href {\doibase 10.1103/PhysRevB.95.235150}
  {\bibfield  {journal} {\bibinfo  {journal} {Phys. Rev. B}\ }\textbf {\bibinfo
  {volume} {95}},\ \bibinfo {pages} {235150} (\bibinfo {year}
  {2017})}\BibitemShut {NoStop}%
\bibitem [{\citenamefont {Dai}\ \emph {et~al.}(2021)\citenamefont {Dai},
  \citenamefont {Schankler}, \citenamefont {Gao}, \citenamefont {Tan},\ and\
  \citenamefont {Rappe}}]{Dai2021}%
  \BibitemOpen
  \bibfield  {author} {\bibinfo {author} {\bibfnamefont {Z.}~\bibnamefont
  {Dai}}, \bibinfo {author} {\bibfnamefont {A.~M.}\ \bibnamefont {Schankler}},
  \bibinfo {author} {\bibfnamefont {L.}~\bibnamefont {Gao}}, \bibinfo {author}
  {\bibfnamefont {L.~Z.}\ \bibnamefont {Tan}}, \ and\ \bibinfo {author}
  {\bibfnamefont {A.~M.}\ \bibnamefont {Rappe}},\ }\href {\doibase
  10.1103/PhysRevLett.126.177403} {\bibfield  {journal} {\bibinfo  {journal}
  {Physical Review Letters}\ }\textbf {\bibinfo {volume} {126}},\ \bibinfo
  {pages} {177403} (\bibinfo {year} {2021})},\ \bibinfo {note} {publisher:
  American Physical Society}\BibitemShut {NoStop}%
\bibitem [{Nak(2017)}]{Nakamura2017}%
  \BibitemOpen
  \href {\doibase 10.1038/s41467-017-00250-y} {\bibfield  {journal} {\bibinfo
  {journal} {Nature Communications}\ }\textbf {\bibinfo {volume} {8}},\
  \bibinfo {pages} {281} (\bibinfo {year} {2017})}\BibitemShut {NoStop}%
\bibitem [{\citenamefont {{Fehske}}\ \emph {et~al.}(2003)\citenamefont
  {{Fehske}}, \citenamefont {{Kampf}}, \citenamefont {{Sekania}},\ and\
  \citenamefont {{Wellein}}}]{Fehske2003}%
  \BibitemOpen
  \bibfield  {author} {\bibinfo {author} {\bibfnamefont {H.}~\bibnamefont
  {{Fehske}}}, \bibinfo {author} {\bibfnamefont {A.~P.}\ \bibnamefont
  {{Kampf}}}, \bibinfo {author} {\bibfnamefont {M.}~\bibnamefont {{Sekania}}},
  \ and\ \bibinfo {author} {\bibfnamefont {G.}~\bibnamefont {{Wellein}}},\
  }\href {\doibase 10.1140/epjb/e2003-00002-2} {\bibfield  {journal} {\bibinfo
  {journal} {European Physical Journal B}\ }\textbf {\bibinfo {volume} {31}},\
  \bibinfo {pages} {11} (\bibinfo {year} {2003})},\ \Eprint
  {http://arxiv.org/abs/cond-mat/0203616} {arXiv:cond-mat/0203616
  [cond-mat.str-el]} \BibitemShut {NoStop}%
\bibitem [{\citenamefont {Rice}\ and\ \citenamefont {Mele}(1982)}]{Rice1982}%
  \BibitemOpen
  \bibfield  {author} {\bibinfo {author} {\bibfnamefont {M.~J.}\ \bibnamefont
  {Rice}}\ and\ \bibinfo {author} {\bibfnamefont {E.~J.}\ \bibnamefont
  {Mele}},\ }\href {\doibase 10.1103/PhysRevLett.49.1455} {\bibfield  {journal}
  {\bibinfo  {journal} {Phys. Rev. Lett.}\ }\textbf {\bibinfo {volume} {49}},\
  \bibinfo {pages} {1455} (\bibinfo {year} {1982})}\BibitemShut {NoStop}%
\bibitem [{\citenamefont {Mildner}\ \emph {et~al.}(2015)\citenamefont
  {Mildner}, \citenamefont {Hoffmann}, \citenamefont {Bl\"ochl}, \citenamefont
  {Techert},\ and\ \citenamefont {Jooss}}]{Mildner2015}%
  \BibitemOpen
  \bibfield  {author} {\bibinfo {author} {\bibfnamefont {S.}~\bibnamefont
  {Mildner}}, \bibinfo {author} {\bibfnamefont {J.}~\bibnamefont {Hoffmann}},
  \bibinfo {author} {\bibfnamefont {P.~E.}\ \bibnamefont {Bl\"ochl}}, \bibinfo
  {author} {\bibfnamefont {S.}~\bibnamefont {Techert}}, \ and\ \bibinfo
  {author} {\bibfnamefont {C.}~\bibnamefont {Jooss}},\ }\href {\doibase
  10.1103/PhysRevB.92.035145} {\bibfield  {journal} {\bibinfo  {journal} {Phys.
  Rev. B}\ }\textbf {\bibinfo {volume} {92}},\ \bibinfo {pages} {035145}
  (\bibinfo {year} {2015})}\BibitemShut {NoStop}%
\bibitem [{\citenamefont {Peierls}(1933)}]{Peierls1933}%
  \BibitemOpen
  \bibfield  {author} {\bibinfo {author} {\bibfnamefont {R.}~\bibnamefont
  {Peierls}},\ }\href {\doibase 10.1007/BF01342591} {\bibfield  {journal}
  {\bibinfo  {journal} {Zeitschrift f{\"{u}}r Physik}\ }\textbf {\bibinfo
  {volume} {80}},\ \bibinfo {pages} {763} (\bibinfo {year} {1933})}\BibitemShut
  {NoStop}%
\bibitem [{\citenamefont {{Dai}}\ \emph {et~al.}(2020)\citenamefont {{Dai}},
  \citenamefont {{Schankler}}, \citenamefont {{Gao}}, \citenamefont {{Tan}},\
  and\ \citenamefont {{Rappe}}}]{Dai2020}%
  \BibitemOpen
  \bibfield  {author} {\bibinfo {author} {\bibfnamefont {Z.}~\bibnamefont
  {{Dai}}}, \bibinfo {author} {\bibfnamefont {A.~M.}\ \bibnamefont
  {{Schankler}}}, \bibinfo {author} {\bibfnamefont {L.}~\bibnamefont {{Gao}}},
  \bibinfo {author} {\bibfnamefont {L.~Z.}\ \bibnamefont {{Tan}}}, \ and\
  \bibinfo {author} {\bibfnamefont {A.~M.}\ \bibnamefont {{Rappe}}},\
  }\href@noop {} {\bibfield  {journal} {\bibinfo  {journal} {arXiv e-prints}\
  ,\ \bibinfo {eid} {arXiv:2007.00537}} (\bibinfo {year} {2020})},\ \Eprint
  {http://arxiv.org/abs/2007.00537} {arXiv:2007.00537 [cond-mat.mtrl-sci]}
  \BibitemShut {NoStop}%
\end{thebibliography}%

\newpage
\end{document}